\documentclass[twocolumn,twoside,slac_two]{revtex4}
\usepackage{graphicx}
\usepackage{fancyhdr}
\begin{document}

\title{MICE Overview}

%

\author{L. Coney (for the MICE Collaboration)}
\affiliation{Department of Physics and Astronomy, University of California-Riverside, Riverside, CA 92521, USA}

\begin{abstract}
Muon ionization cooling provides the only practical solution for preparing high brightness beams necessary for
a neutrino factory or muon collider.  The Muon Ionization Cooling Experiment (MICE) is under development at
the Rutherford Appleton Laboratory (UK).  It comprises a dedicated beam line designed to generate a range of
input emittances and momenta with time-of-flight and Cherenkov detectors to select a pure muon beam.  A first
measurement of emittance is performed in the upstream magnetic spectrometer with a scintillating fiber
tracker.  A cooling cell will then follow, alternating energy loss in liquid hydrogen and acceleration by RF
cavities.  A second spectrometer identical to the first and another particle identification system provide a
measurement of the outgoing emittance.  In late 2009, it is expected that the beam and many of the particle
identification detectors will be in the final commissioning phase, and the first measurement of input beam
emittance will take place in 2010.  The steps of commissioning, emittance measurement and cooling will be
described. 
\end{abstract}

\maketitle

\thispagestyle{fancy}


\section{Introduction}
\begin{figure}[h]
\centering
\includegraphics[width=80mm]{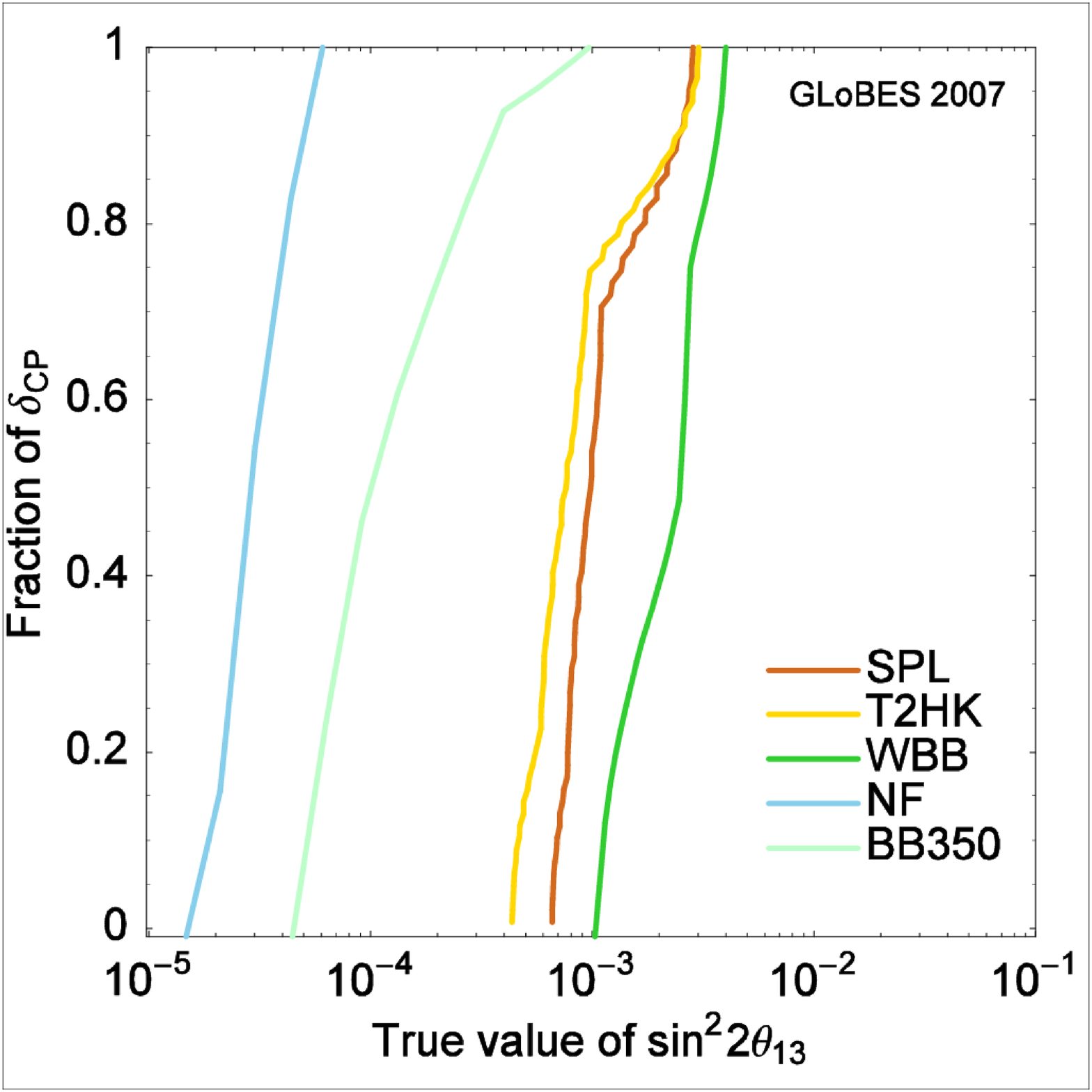}
\caption{Plot modified from that in the ISS Physics Working Group Summary Report~\cite{NuFact_physics1}, to show only the best-case 
scenario for each facility, with 3$\sigma$ curves showing the discovery
reach of several facilities in $\sin^{2}2\theta_{13}$. The limits are shown as a function of $\sin^{2}2\theta_{13}$ (x-axis) and of the fraction of all
possible values of the CP phase (y-axis). In orange is the reach of the SPL super-beam, in yellow is that of T2HK, in green
is that of the wide-band beam experiment, and in light green is that of the beta-beam.  With a significantly better sensitivity limit of 
$\sin^{2} 2\theta_{13}$ $\sim$ $1.5 \times 10^{-5}$, the 
discovery reach of the Neutrino Factory is shown in blue.} \label{ISS_Theta_13}
\end{figure}

\begin{figure}[h]
\centering
\includegraphics[width=80mm]{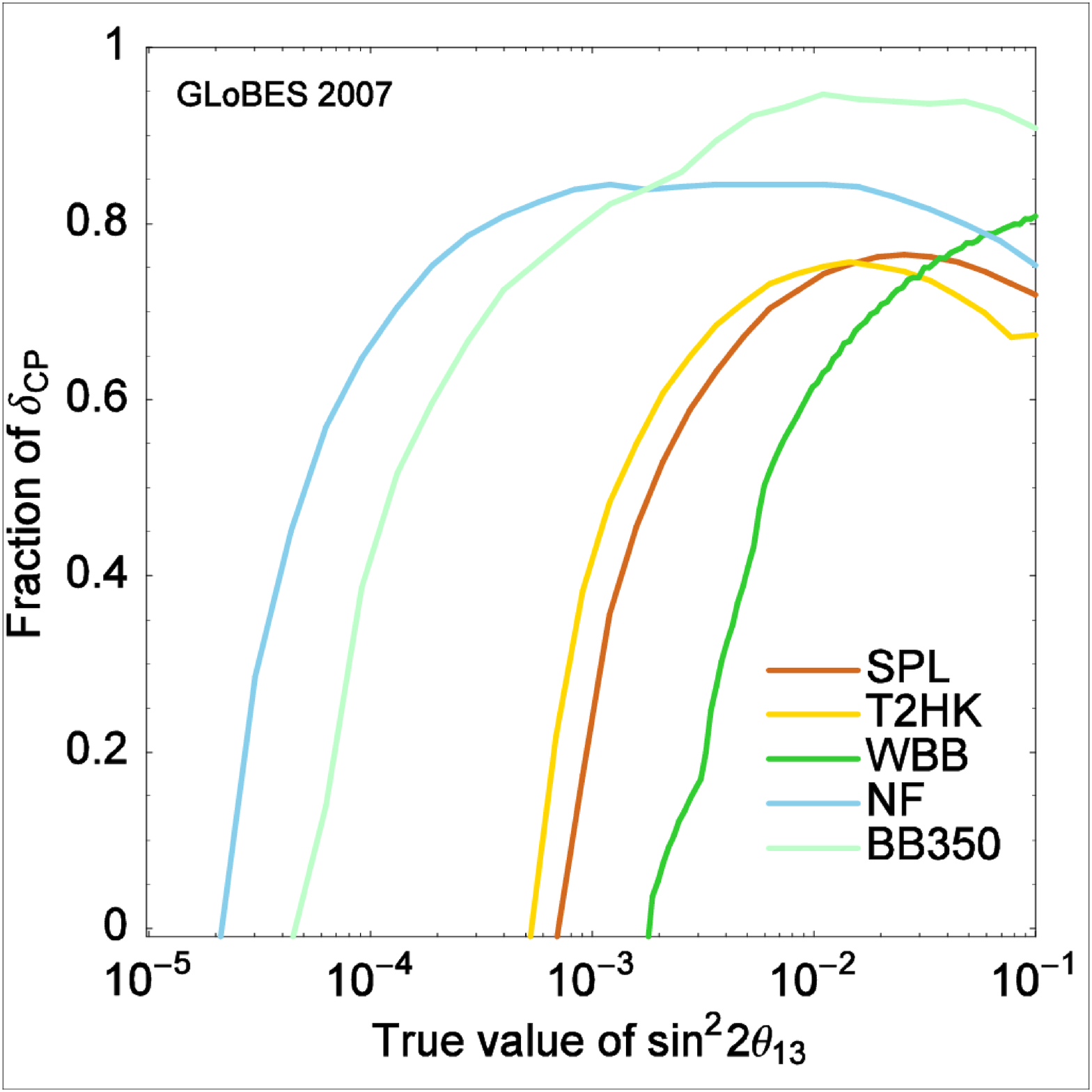}
\caption{Plot modified from that in the ISS Physics Working Group Summary Report~\cite{NuFact_physics1}, to show only the best-case 
scenario for each facility, with 3$\sigma$ curves showing the discovery
reach of several facilities in the CP phase $\delta$.  The limits are shown as a function of $\sin^{2}2\theta_{13}$ (x-axis) and of the fraction of all
possible values of the CP phase (y-axis). In orange is the reach of the SPL super-beam, in yellow is that of T2HK, in green
is that of the wide-band beam experiment, and in light green is that of the beta-beam. The discovery reach of the Neutrino Factory 
is shown in blue, and for values of $\sin^{2} 2\theta_{13} \le 4 \times 10^{-3}$ it has better coverage than all other options shown.} \label{ISS_CP}
\end{figure}

The next generation particle physics facility may be a Neutrino Factory~\cite{NuFactory}, to advance in-depth study of the neutrino
sector, or a Muon Collider~\cite{MC}, for precision Higgs physics or as an energy frontier machine. 
Each of these facilities has been proposed to furthur the understanding of fundamental questions in
particle physics.  In a Neutrino Factory, the decay of stored muons ($\mu^-$) produces a well-understood, fully
characterized, intense and narrow beam of muon neutrinos ($\nu_{\mu}$) and electron anti-neutrinos ($\bar{\nu}_e$). 
This allows the study of the Golden oscillation channel: $\nu_e \rightarrow \nu_{\mu} $ and  $\bar{\nu}_e \rightarrow
\bar{\nu}_{\mu} $, where the sign of the detected muon is opposite that in the
storage ring.  This uniquely clean source of neutrinos can be used to study neutrino oscillations and leptonic CP
violation.  With $10^{21}$ muon decays per year, these key processes would enable the study of the neutrino mass
hierarchy, measurement of $\sin^{2}2\theta_{13}$ to ~$10^{-4}$ (see ~Figure~\ref{ISS_Theta_13}), and would provide the best 
chance of discovering CP violation in the lepton sector (see
~Figure~\ref{ISS_CP})~\cite{NuFactory,NuFact_physics1,NuFact_physics2}.

A Muon Collider as the next generation lepton collider ($\mu^+ \mu^-$) would provide a facility to perform low energy
(100-1000 GeV) precision studies of the Higgs boson in the Standard Model or in possible extensions such as supersymmetry. A Muon
Collider would also allow exploration of center-of-mass
energies up to 4 TeV~\cite{MC_physics1,MC_physics2,MC_physics3}. Because muons are much more massive than electrons,
they have negligible synchrotron radiation, thus making possible a high-energy circular lepton collider with a very small
footprint. In addition, $\mu^+ \mu^-$ collisions have much less beamstralung radiation and are therefore more
monochromatic than collisions in $e^+ e^-$ machines. With such advantages over traditional $e^+ e^-$ colliders,
consideration must be given to the possibility of creating a Muon Collider with guidance regarding energy scale to come
from Large Hadron Collider (LHC) results.

In both the Neutrino Factory and Muon Collider, the muon beam is
produced from the decay of secondary pions generated by proton collisions with a target. Due to this manner of
production, the initial muon beam has a very large energy spread and large spatial dimensions. Because both of these
accelerators require high brightness muon beams to reach the intended physics goals, cooling of the initial muon beam
is central to the generation of the required muon intensity.

With such a short particle lifetime of 2.2 $\mu$s, traditional beam cooling techniques, like stochastic cooling, 
cannot be used for a muon
beam. Ionization cooling provides an alternative method to quickly reduce the transverse momentum spread of the beam
and increase the fraction of muons fed into the acceleration portion of the Neutrino Factory or Muon Collider. In ionization cooling, 
the muon beam passes through liquid hydrogen (L$H_2$) absorbers followed by accelerating
RF cavities. The beam loses both longitudinal and transverse momentum in the absorbers, and acceleration in the RF
cavities restores only the longitudinal component of the momentum.  In this manner, the transverse emittance of the
muon beam is reduced quickly and the beam is cooled. 

MICE~\cite{MICE_proposal} is an experimental program whose purpose is to build, commission, and test a fully 
engineered section of an ionization cooling channel based on a design from the Feasibility Study-II~\cite{FeasibilityII}.  
This cooling
channel will be long enough to cause a 140-240 MeV/c muon beam to undergo a 10$\%$ reduction in transverse emittance, to be measured with
a precision of 1$\%$. A lead diffuser of variable thickness will be used to tune the incoming transverse beam emittance 
from 2 to 10 $\pi$ mm-rad. Particle identification detectors ensure a pure muon beam, and identical scintillating 
fiber spectrometers will precisely measure the beam emittance both before and after the cooling channel. MICE will 
provide the first demonstration of ionization cooling and will offer insight into the practical challenges involved 
in building a muon accelerator. MICE is an international collaboration consisting of a unique blend of more than 100 
particle and accelerator physicists and engineers from the US, the UK, Switzerland, the Netherlands, Japan, Italy,
China, Bulgaria, and Belgium contributing to build this unusual experiment and advance our capability
to create the next generation lepton accelerator. 

\section{MICE Overview}
\begin{figure}[h]
\centering
\includegraphics[width=80mm]{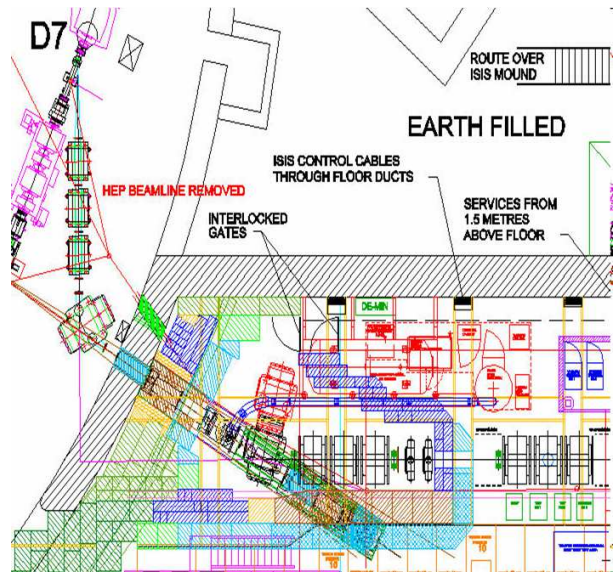}
\caption{MICE beamline branching off of the ISIS proton synchrotron at RAL in the UK. The MICE target sits in sector 7
of the ISIS ring.} \label{MICE_beamline}
\end{figure}

The MICE experiment is under construction at Rutherford Appleton Laboratory (RAL) on a dedicated muon beamline at the
ISIS 800 MeV proton synchrotron (see ~Figure~\ref{MICE_beamline}).  A hollow cylindrical titanium target, designed 
and built by
collaborators from the UK, is dipped into the proton beam at the end of the ISIS 20 ms beam cycle. Because MICE is not 
the sole active experiment at the accelerator, the target runs parasitically during standard ISIS user periods,   
dipping once every 50 ISIS cycles at a rate of $\sim$0.4 Hz.

\begin{figure*}[t]
\centering
\includegraphics[width=135mm]{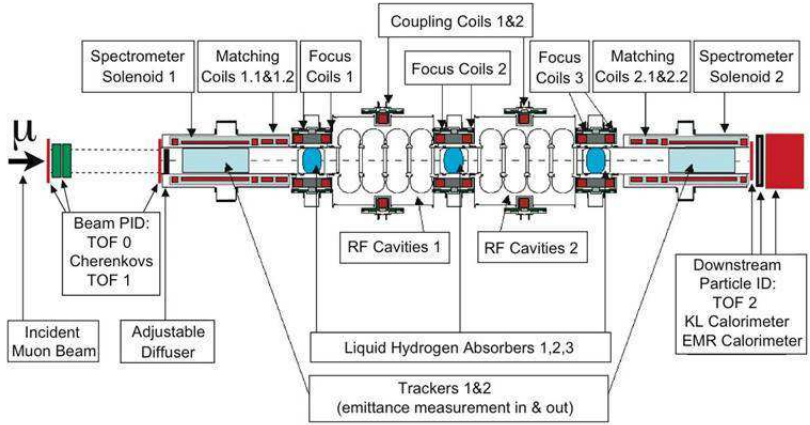}
\caption{Schematic of the complete MICE cooling channel including the incoming and outgoing particle identification detectors,
the two scintillating fiber spectrometers, three low-Z liquid hydrogen absorbers, and accelerating RF cavities.} \label{MICEcartoon}
\end{figure*}

Pions from interactions in the target are captured by a quadrupole triplet within the synchrotron enclosure and
then momentum-selected by the first MICE dipole magnet. Just after this dipole is a 5-m-long, 12-cm-bore, 5-T 
superconducting decay solenoid which contains both pions coming from the target and muons from pion decay.  A second 
dipole magnet then
selects a pure beam of muons by specifically bending particles with lower momentum than at the first dipole.  The muon
beam is then transported by a quadrupole triplet through particle identification detectors (PID). Two aerogel
threshold Cherenkov counters (US-Belgium) and a set of time-of-flight (TOF) detectors (Italy-Bulgaria) together provide
excellent $\pi$/$\mu$ separation up to 300 MeV/c, and muon beam purity is ensured to better than
99.9$\%$. Finally, the last large acceptance quadrupole triplet transports the beam into the MICE cooling channel (see
~Figure~\ref{MICEcartoon}).

At each end of the cooling channel are TOF detectors and two 1.1-m-long, 20-cm-radius scintillating fiber trackers
(UK-US-Japan),each with five position measurement stations. These spectrometers are located inside 2-m-long, 4-T,
uniform-field
superconducting solenoid magnets (US).  These complex spectrometer magnets also have four additional coils: two to match
optics with the cooling channel, and two end coils to ensure field uniformity. The incoming and outgoing 6D muon beam
emittances are measured by determining particle momenta and spatial coordinates with the trackers and measuring time with
the TOF detectors. This time measurement will also be used to determine the phase of the electric field seen by the
the muons during passage through the RF cavities.

\begin{figure}[h]
\centering
\includegraphics[width=80mm]{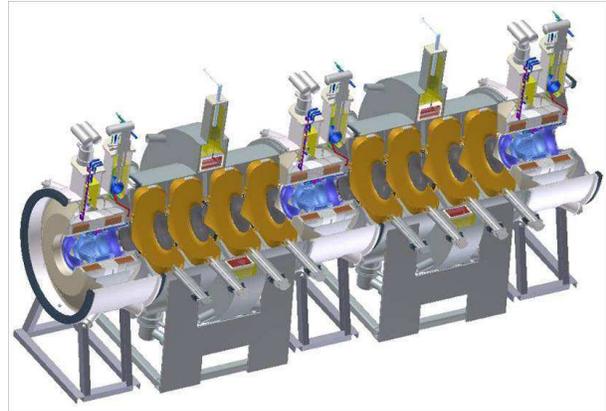}
\caption{Engineering drawimg of the complete MICE cooling channel including two RF-Coupling Coil (RFCC) modules and three 
Absorber-Focus-Coil (AFC) modules.} \label{3D_cooling}
\end{figure}

The cooling channel itself is made up of alternating low-Z liquid hydrogen absorbers and normal-conducting 201-MHz RF 
cavities.  The $LH_2$ absorbers are each situated within an Absorber-Focus-Coil module (AFC) consisting of 
superconducting coils designed
to provide strong focusing at the absorbers. This focusing maximizes the effectiveness of the ionization cooling.  The
RF cavities must be low frequency in order to accommodate the large beam size. They are normal-conducting because
they must operate within a focusing magnetic field which presents many technical challenges.  The complete cooling 
channel is made up of two RF-Coupling-Coil (RFCC) modules and three AFC modules (see~Figure~\ref{3D_cooling}).

In addition to the final TOF detector at the downstream end of the cooling channel, downstream PID is done with a
calorimeter (Italy-Geneva-Fermilab) designed to distinguish between muons and electrons from muons that have decayed 
during flight through the experiment.  The first part of the calorimeter is the KLOE-Like (KL) lead-scintillating-fiber
sandwich layer (Italy) which degrades electrons.  This will be followed by the Electron-Muon Ranger (EMR), a 1-$m^3$
block of extruded scintillator bars.  This final detector at the downstream end of MICE is designed to measure muon 
momentum by range.

\section{MICE Status}
\subsection{Beamline and Target}
\begin{figure}[h]
\centering
\includegraphics[width=80mm]{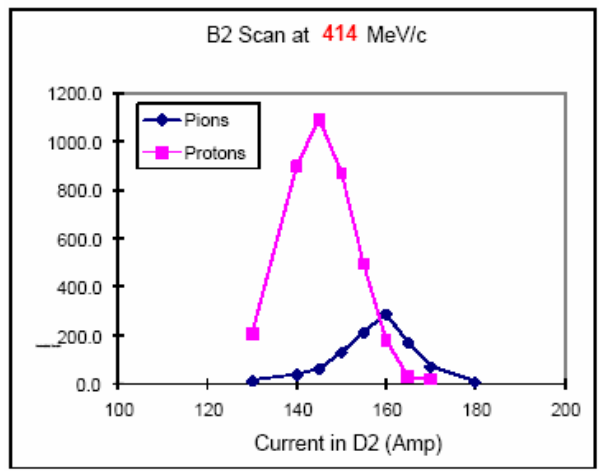}
\caption{Scan of current in the second MICE bending dipole (D2) to understand the effect on particle type and rate through the
downstream section of the MICE beam line. Shown are the relative proton and pion rates as a function of current in D2 for
particle momentum of 414 MeV/c selected by the first bending magnet~\cite{JSG_PAC}.} \label{D2_scan}
\end{figure}

Commissioning of the MICE beamline began in fall of 2008 with data-taking to understand the conventional beamline magnet
performance~\cite{JSG_PAC}.  Protons, pions, positrons, and muons were clearly seen and magnet settings for different particle
types and momenta were determined.  Scans were done using the two bending magnets to maximize particle rate in
scintillation counters along the beamline (see~Figure~\ref{D2_scan}). These early commissioning data were taken without the 
5-T decay solenoid from PSI.  Missing multi-layer insulation made it impossible for all coils of the magnet to become
superconducting.  However, in April of 2009, this problem was corrected and the magnet was successfully ramped up to its full operating current.
Since then, the decay solenoid has been commissioned and is now operated routinely as part of the MICE beamline.

\begin{figure*}[t]
\centering
\includegraphics[width=135mm]{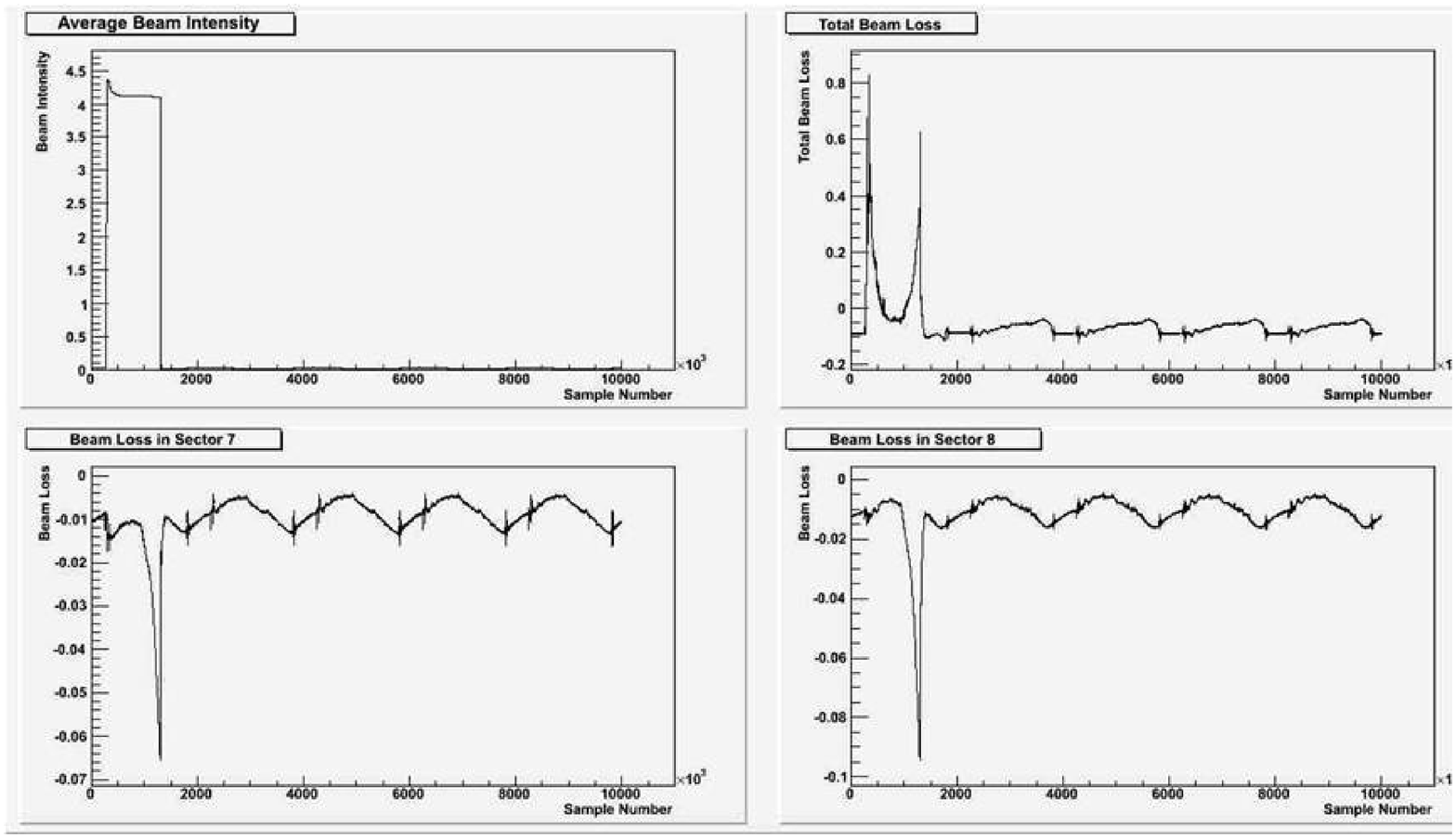}
\caption{ISIS accelerator information available during MICE running for a dedicated MICE study period. During regular running,
all five cycles shown would have ISIS beam and the corresponding beam losses. The top left plot 
shows ISIS beam intensity, the top right shows total beam loss for a single cycle. The bottom left and right plots show beam 
loss due to the MICE target in sectors 7 and 8 of ISIS.} \label{TargetDAQ}
\end{figure*}

Data-taking recently resumed in September of 2009 with some dramatic changes in the MICE beam line. As previously mentioned, the 5-T
decay solenoid is now operational, and has significantly increased the particle rate in MICE.  Studies are currently being done to
quantify the effects of the solenoid for different beamline settings.  The MICE target has also been redesigned and rebuilt since a
target failure was suffered in December of 2008.  The original target ran for 190k dips into the ISIS beam before becoming jammed
due to a problem with a loose alignment collar.  In the following eight months, the target was re-engineered with
a simpler hollow circular shaft, no alignment collar, and better tolerances on the bearings.  The new target was installed in the
ISIS beamline in August 2009 and was used in the first data-taking run of the year in September. The target has operated very well
and was dipped over 60,000 times into the beam without any problems.
During the September run, several studies were done to determine the optimal target operating settings.  Target dip timing with
respect to the ISIS cycle was varied as well as target dip depth into the beam. These parameters were modified in order to maximize 
particle production rate in the MICE beam line and to intersect the correct energy of particles in ISIS. Tests were also done to
ensure that MICE would not adversely affect normal ISIS operations during parasitic running. To better understand the
interaction between the MICE target and the ISIS beam, live beam loss information is fed into the MICE control room during running.
We are able to monitor the ISIS beam intensity, the total beam loss during each cycle, and the beam loss in sectors 7 and 8 of the
ISIS ring (see~Figure~\ref{TargetDAQ}). These are the areas closest to the target, and beam loss levels in these regions determine MICE running conditions. 
Studies are ongoing to quantify MICE particle production rates as a function of loss levels in ISIS for both positive and negative
beam particles.  

\subsection{Particle Identification Detectors}
\begin{figure}[h]
\centering
\includegraphics[width=80mm]{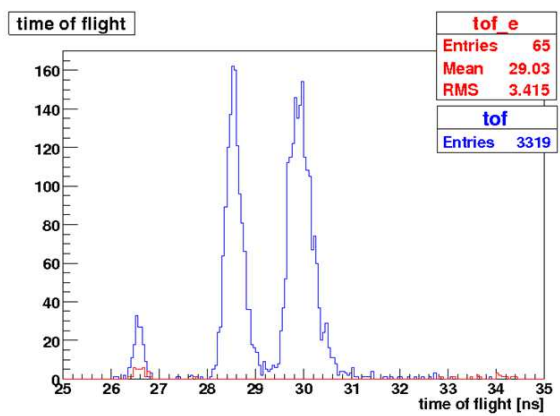}
\caption{Time-of-flight distribution showing $e^+$, $\mu^+$, and $\pi^+$ particles in a nominal 300 MeV/c pion beam.} \label{TOF}
\end{figure}

Commissioning of the first set of TOF detectors in the MICE beamline was begun in late 2008. The TOF0 and TOF1 detectors are 
located after the second and third set of quadrupoles, respectively, and are separated by about 10 meters.  They are made 
up of X and Y counter arrays of inch-thick scintillator bars with dual photomultiplier (PMT) readout, with approximately 50 cm 
by 50 cm active area.  Using a set of pion data runs, they were found to have timing resolutions of 55 to 65 ps, which was very
close to design specification~\cite{TOF_PAC}. Figure~\ref{TOF} shows the time-of-flight distribution for particles in a beamline 
designed to select 300 MeV/c pions. Good separation can be seen between positrons, muons, and pions in this early analysis result.  The
third TOF detector, which will ultimately be located after the full cooling channel, has been fabricated and will be delivered to RAL for
installation in the beam line in December 2009.

\begin{figure}[h]
\centering
\includegraphics[width=80mm]{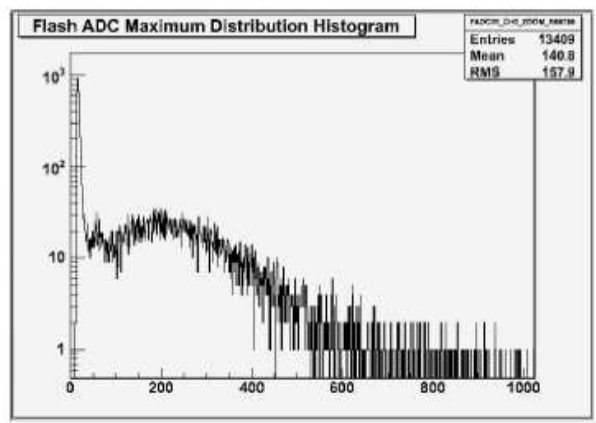}
\caption{Light spectrum from a single PMT on one of the aerogel threshold Cherenkov detectors.} \label{CKOV}
\end{figure}

Two high density aerogel threshold Cherenkov (CKOV) detectors~\cite{TOF_PAC} designed to enable $\pi$/$\mu$ separation at higher momenta, above 300 
MeV/c, are located in the MICE beamline just downstream of the first TOF detector. By using the TOF system together with the
threshold Cherenkov detectors, 4-5$\sigma$ $\pi$/$\mu$ separation can be achieved over the full range of particle momenta expected
in MICE.  The two detectors have different indices of refraction with corresponding muon threshold momenta of 278
MeV/c and 220 MeV/c, respectively.  The Cherenkov light from each counter is read out by four PMTs, and a typical
PMT spectrum can be seen in Figure~\ref{CKOV}.  Approximately 20-25 photoelectrons per counter are collected, which allows for a
muon tagging efficiency of 98$\%$ over the 220-360 MeV/c momentum range of particles in MICE. Commissioning of these detectors 
was begun in 2008 and has continued during the latest run periods.

The KL lead-scintillating-fiber sandwich layer portion of the calorimeter system~\cite{TOF_PAC} is located at the far 
downstream end of the MICE beamline. It is designed to degrade electrons in the beam and has 4 cm active depth, corresponding 
to 2.5 $X_0$.  The KL detector was installed in the beamline, commissioned with cosmic rays, and exposed to pion and
positron beams in 2008.  This process continued in September 2009 with dedicated data-taking runs of 150 MeV/c and 300 MeV/c 
positrons and electrons focused onto the KL detector. Construction has also begun on the final calorimeter, the EMR, which is being
built at the University of Geneva, and delivery of the detector is planned for July 2010.

\subsection{Spectrometers}
\begin{figure}[h]
\centering
\includegraphics[width=80mm]{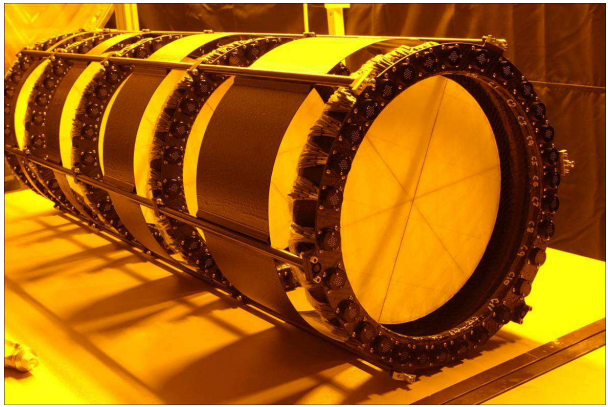}
\caption{Completed scintillating fiber tracker showing five measurement stations, each with three planes of fiber doublets rotated 120
degrees with respect to each other.} \label{Tracker}
\end{figure}

\begin{figure}[h]
\centering
\includegraphics[width=80mm]{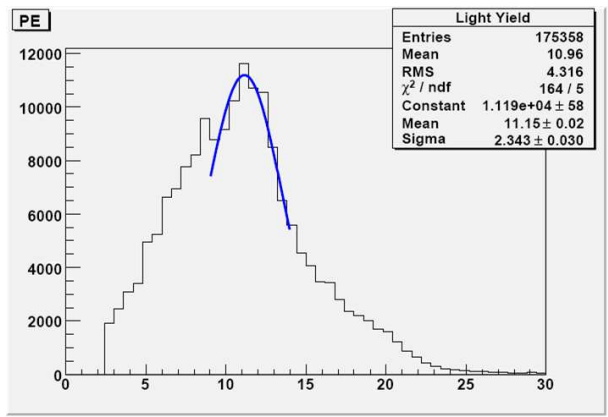}
\caption{Light yield from clusters used in tracks for the first tracker module. Measured using $\sim$175,000 hits corresponding to $\sim$10,000 cosmic 
ray tracks.} \label{trackerLight}
\end{figure}

In MICE, spectrometers using particle physics techniques will measure the 10$\%$ reduction in transverse beam emittance with an
absolute precision of 0.1$\%$.  This will be done using two scintillating fiber tracking detectors inside 
solenoid magnets on each
side of the cooling channel~\cite{Tracker_PAC}.  Each detector has five measurement stations with three planes of 350 $\mu$m fiber doublets
rotated 120 degrees with respect to each other, thus allowing reconstruction of space points along the path of the muons.  These 
two tracking detectors are completed (see~Figure~\ref{Tracker}), and both have been tested using cosmic rays.  Cosmic
ray testing of the first tracker module is finished, and the detector performed as designed. Figure~\ref{trackerLight} shows the distribution of light from
clusters used in tracks. This measurement was made using $\sim$175,000 hits corresponding to roughly 10,000 cosmic ray tracks.  The
average light yield of 11 photoelectrons met the design goal of 10.5 photoelectrons. While this is a small amount of light, the fibers
are read out using Visible Light Photon Counters~\cite{VLPC}, solid state photo-detection devices with high quantum efficiency
(QE) of $\sim$80$\%$, high gain ($\sim$50,000), and low noise.  Having met the design goal, we can be confident that, in spite of the
very small diameter fibers chosen in order to minimize the amount of material traversed by muons, the tracker will be able to
reconstruct particle tracks with high efficiency. This tracker also has fewer 
than 0.1$\%$ dead channels and a position resolution consistent with the design goal of 430 $\mu$m.

The two 4-T spectrometer solenoids~\cite{Spectrometer_PAC} that will house the scintillating fiber trackers are being fabricated by Wang NMR, 
Inc., in conjunction
with engineers from the Lawrence Berkeley National Laboratory (LBNL).  Final assembly of the first spectrometer solenoid was completed, the
magnet was cooled down to superconducting temperature, and training of the coils was begun. However, the magnet did not pass the
acceptance tests, and modifications to the design are now being made.  After the changes are made to the magnet, it will be retested
and then shipped to the Fermi National Accelerator Lab (FNAL) for magnetic field measurements in late 2009.  After field testing, it will
be shipped to RAL where the TOF, diffuser, tracker, and magnet will be integrated and the full spectrometer unit will be installed in
the MICE beam line. The second spectrometer solenoid will be delivered three months after the first.

\subsection{Cooling Channel}
In the cooling channel, the liquid-hydrogen absorbers reduce both transverse and longitudinal particle momentum.  Hydrogen was chosen
because it causes the least multiple scattering, or beam heating, for the amount of momentum absorbed compared to any other element.  
The first internal convection cooling $LH_2$ absorber has been fabricated at KEK (Japan) and is undergoing thermal testing. Ten
of the twenty thin aluminum windows for the absorbers have been made at the University of Mississippi.  These 30-cm-diameter windows are 
curved for added strength and tapered to 180 $\mu$m at the center to minimize muon scattering, which would counteract the desired
cooling. Each of the $LH_2$ absorbers sits inside an Absorber Focus Coil (AFC).  The AFCs are being manufactured in the UK with delivery
expected in 2010.

\begin{figure}[h]
\centering
\includegraphics[width=80mm]{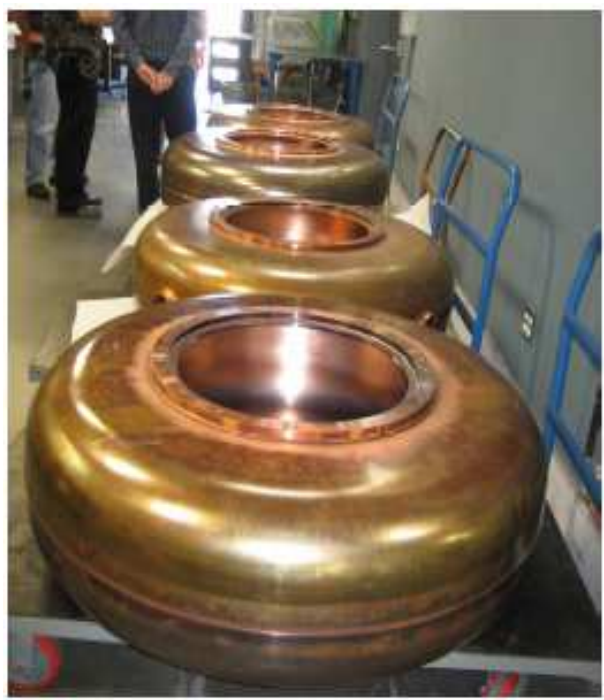}
\caption{The first set of four copper 201 MHz RF cavity bodies during fabrication.} \label{RFcavities}
\end{figure}

The rest of the cooling channel is made up of accelerating RF cavities. The final 
design review of the MICE RF Coupling Coil (RFCC) modules, consisting of four normal conducting 201-MHz cavities with a guiding magnetic
field provided by a large diameter coupling coil, was finished in 2008.  The coupling coils were designed by the Harbin
Institute of Technology in China who are also responsible for oversight of their fabrication. This process has begun, and the prototype
coupling coil will be delivered for use in the MUCOOL test area at Fermilab in late 2010, followed by the first complete MICE coupling
coil delivery in summer 2011. LBNL is responsible for the fabrication of the RF cavities and integration with the RFCC modules for 
delivery to RAL in late 2011. The first five copper cavities are nearly complete (see~Figure~\ref{RFcavities}). Each cavity also requires two TiN coated beryllium windows with a specially designed curve to prevent buckling due to
thermal expansion during RF heating.  The first two windows have been delivered to LBNL and are undergoing quality control testing.  

Once delivered to RAL and installed in the MICE beamline, each MICE cavity will require $\sim$1 MW of RF power in a 1-ms pulse at a rate of 1
Hz.  This is provided by four sets of amplifiers including Burle 4616 drive amps and 2.5 MW Thales TH116 main amplifiers donated by LBNL and
CERN.  Significant refurbishment of the amplifiers is being done at Daresbury Laboratory (UK), where the RF power system will also be 
tested~\cite{RF_PAC}. The first 4616 amplifier has been refurbished and preliminary tests have been done.  The low level RF system to drive
this unit has been tested successfully against a dummy load and $\sim$2 kW of power was achieved. The first TH116 has been cleaned, 
repaired, the RF surfaces have been silver plated, and it is now ready for testing. A plan to test the 4616 system in steps up to 1 MW will
be completed by the end of 2009.

\section{MICE Schedule}
\begin{figure}[h]
\centering
\includegraphics[width=80mm]{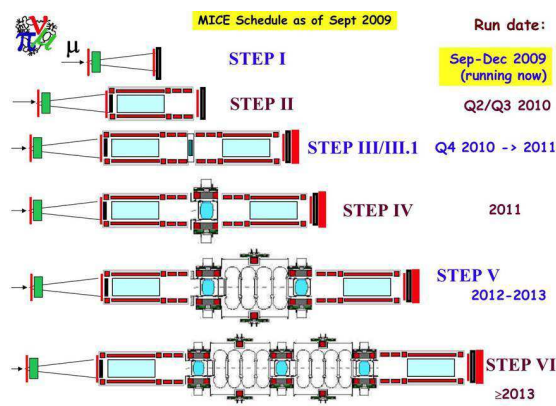}
\caption{The MICE steps and the schedule for completing each step.} \label{schedule}
\end{figure}

MICE is proceeding in a staged manner (see~Figure~\ref{schedule}) to ensure an accurate understanding of the detectors, beamline, and
measurement techniques, and to meet funding
profiles.  The beamline, which delivers muons to the cooling channel, and the experimental hall infrastructure for the first three
steps are completed.  Step I, commissioning of the beamline and PID detectors, began in 2008 and has continued during the September 2009 ISIS
User Run.  With a new target installed and the decay solenoid completely operational, measurement of particle production in MICE as a
function of beam loss in ISIS is possible, and characterization of the muon beam will begin. Calibration of beamline simulations can be done
in detail with the beamline fully operational, and optimization of magnet settings will be done.  Commissioning of the TOF system, the CKOV
detectors, and the KL calorimeter will also be completed in Step I, which will run through early 2010. In summer of 2010, for Step II, the
first tracker and spectrometer solenoid module will be installed in the MICE beamline.  When this happens, it will be possible to make the first
precision measurement of the incoming MICE muon beam emittance.

The second spectrometer solenoid will follow three months after the first, and will be installed in MICE in the summer of 2010 for the start
of Step III.  During 
this step, the muon beam emittance will be measured in both spectrometers, and the tracking detectors can be fully tested, compared,
and calibrated.  This comparison, without any expected change in emittance, will allow the determination of any measurement biases 
present and will test correction procedures. In Step III.I, a solid LiH disk will be placed between the two spectrometers, and will provide
an opportunity to test muon interaction in a different absorber material. While hydrogen possesses ideal properties for cooling, the $LH_2$
system brings with it many challenges including satisfying safety requirements, the use of very thin containment windows, and storage of the
hydrogen.  Because of these features, MICE has developed the capability to test other absorber materials.  Another potential modification 
to Step III involves the addition of a plastic wedge absorber between the spectrometers to investigate emittance exchange.  Step III should 
run by the end of 2010, and with its completion, the MICE beamline dynamics, particle production, PID systems, spectrometers, and
muon beam emittance measurement techniques will be fully understood.

In Step IV, running in 2011, the first $LH_2$ absorber-focus-coil module will be installed and will allow the first measurement of muon
cooling to be made.  Tests of the focusing optics will also be performed during Step IV, by exploring configurations with different
magnetic-field reversals.  Step V will follow in 2012 when the first RF coupling coil module is installed and MICE will perform the first
test of sustainable cooling, where RF cavities restore the longitudinal momentum lost in the absorbers.  

Finally, Step VI, running in 2013, will meet the primary goal of MICE by operating a complete cooling channel with three $LH_2$ absorber
modules and two RF coupling coil modules.  Several configurations of the focusing optics in the central absorber will be fully tested,
comparisons with Monte Carlo simulations will be made, and a detailed understanding of ionization cooling will be demonstrated.  Information
from the MICE studies will guide the design of future muon accelerators, provide a practical understanding of challenges presented by a
realistic ionization cooling channel, and present solutions to these challenges.  




\begin{acknowledgments}
This work is supported by the U.S. National Science Foundation under grant No. PHY-0630052.  The author would like to thank the
UC-Riverside group and the MICE collaboration for their hard work, support, and for the opportunity to present the work being done 
by the collaboration.  

\end{acknowledgments}

\bigskip 

\end{document}